\documentclass[12pt]{iopart}
\usepackage{iopams}

\usepackage{dsfont}
\usepackage{color}
\usepackage{placeins}

\usepackage{amssymb,xspace,graphicx,ifthen,amsfonts}
\usepackage[utf8]{inputenc}
\usepackage{amsfonts}
\usepackage{colordvi}
\usepackage{pifont}
\usepackage{appendix}

\newlength{\ldag}
\begin{document}

\title[Effective Spin Couplings in the Honeycomb Mott Insulator]{Effective Spin Couplings in the Mott Insulator of the Honeycomb Lattice Hubbard Model}

\author{Hong-Yu Yang}
\address{Institute of Theoretical Physics, Ecole Polytechnique F\'ed\'erale de Lausanne, CH-1015 Lausanne, Switzerland}

\author{A.~Fabricio Albuquerque}
\address{Instituto de F{\' \i}sica, Universidade Federal do Rio de Janeiro, Cx.P.~68.528, CEP 21941-972 Rio de Janeiro-RJ, Brazil}

\author{Sylvain Capponi}
\address{Laboratoire de Physique Th{\' e}orique, Universit{\' e} de Toulouse and CNRS, UPS (IRSAMC), F-31062 Toulouse, France}

\author{Andreas M. L\"auchli}
\address{Institut f\"ur Theoretische Physik, Universit\"at Innsbruck, A-6020 Innsbruck, Austria}

\author{Kai Phillip Schmidt}
\ead{kai.schmidt@tu-dortmund.de}
\address{Lehrstuhl f\"ur Theoretische Physik I, Otto-Hahn-Stra\ss e 4, TU Dortmund, D-44221 Dortmund, Germany}


\begin{abstract}
Motivated by the recent discovery of a spin liquid phase for the Hubbard model on the honeycomb lattice at half-filling \cite{Meng10}, we apply both perturbative and non-perturbative techniques to derive effective spin Hamiltonians describing the low-energy physics of the Mott-insulating phase of the system. Exact diagonalizations of the so-derived models on small clusters are performed, in order to assess the quality of the effective low-energy
theory in the spin-liquid regime. We show that six-spin interactions on the elementary loop of the honeycomb lattice are the dominant sub-leading effective couplings. A minimal spin model is shown to reproduce most of the energetic properties of the Hubbard model on the honeycomb lattice in its spin-liquid phase. Surprisingly, a more elaborate effective low-energy spin model obtained by a systematic graph expansion rather disagrees beyond a certain point with the numerical results for the Hubbard model at intermediate couplings.   
\end{abstract}


\maketitle

\section{Introduction}

The quest for exotic quantum phases lacking conventional long-range order in dimensions higher than one has recently attracted enormous
interest, for fascinating unusual properties, such as quantum number fractionalization and/or anyonic excitations, are expected to emerge from
certain spin Hamiltonians \cite{Kitaev03,Balents10}. In this context, magnetic systems are an important playground, since exactly solvable
models displaying such properties have been recently introduced in the literature \cite{Kitaev03}.

Common wisdom has it that, if experimental realizations of such exotic phases are to be found, one should look at Hubbard-like systems deep into
their strongly interacting regime, on {\em frustrated} geometries disfavoring more conventional N{\' e}el order. Indeed, in the limit of large on-site
interactions $U$, where charge fluctuations are inhibited, the Hubbard Hamiltonian maps to the Heisenberg model with nearest-neighbor (NN)
interactions only, for which numerical evidence for a gapped quantum spin liquid (QSL) ground-state has been recently found on the highly
frustrated kagom{\'e} lattice \cite{Yan10}. 

An alternative route has been pursued in recent years, by focusing on two-dimensional Hubbard models in the regime of {\it intermediate} $U$, where
charge fluctuations soften the Mott insulating phase. There is indeed convincing evidence that a QSL phase is realized on the frustrated triangular lattice \cite{Morita02,Motrunich05,Kyung06,Sahebsara08,Tocchio08,Yoshioka09,Yang10}. Sizeable charge fluctuations manifest themselves as long-range and/or multi-body
effective spin interactions, beyond NN spin-exchange. In particular, it has been shown that four-spin exchanges are the dominant correction to the
Heisenberg model on the triangular geometry, accounting for the emergence of a QSL in this case \cite{Motrunich05,Yang10}.

Whichever of the aforementioned routes is followed, it appears that frustration is an essential ingredient for a QSL. In face of this common expectation, the
recent discovery \cite{Meng10} of a QSL phase for the Hubbard model on the {\it unfrustrated} honeycomb lattice, at half-filling and intermediate $U$, is
consequently very surprising and has attracted enormous interest. In particular, under the light of our previous discussion, the question of what the effective
spin interactions are for stabilizing a QSL in this case naturally arises.

An effective spin model comprising terms of up to fourth-order in $t/U$ is basically a frustrated $J_1$-$J_2$ model, since four-spin interactions, the dominant
corrections to the NN Heisenberg model on square and triangular geometries \cite{Yang10}, are strongly suppressed on the honeycomb lattice,
for it lacks length-four loops. Due to this fact, it has been argued \cite{Wang2010,Cabra11,Clark11} that the emergence of a QSL can be simply accounted to the
frustrating next-NN coupling $J_2$, a claim that has spurred a number of works on the so-called $J_1$-$J_2$ model \cite{Albuquerque11,Reuther11,Oitmaa11,
Mosadeq11,Mezzacapo12}. Although such studies differ in detail, they commonly detect a quantum critical point, at $J_2 /J_1\approx 0.2$, between long-range
ordered N\'eel and a magnetically disordered phase. The precise nature of this non-magnetic state remains somehow unclear, albeit most numerical results
point to a valence bond solid (VBS) \cite{Albuquerque11,Reuther11,Oitmaa11,Mosadeq11}. However, and interestingly enough, in Ref.~\cite{Mezzacapo12}
a variational approach finds that such VBS becomes unstable towards a gapped spin liquid if large enough length scales are taken into account.

Independently of the nature of magnetically disordered phase stabilized for the $J_1$-$J_2$ model, it is crucial to gauge its validity as a low-energy theory for
the honeycomb lattice Hubbard Hamiltonian in the intermediate range of $t/U$, where the QSL emerges. Indeed, evidence indicating a more involved situation
has been found in Ref.~\cite{Yang11}: the bare series expansion (of up to order $14^{\rm th}$) is {\em not} converged for values of $t/U$ leading to the QSL.
Furthermore, a non-perturbative derivation of effective spin interactions, obtained via graph-based continuous unitary transformations \cite{Yang11}, yields
a relatively small ratio $J_2 /J_1 < 0.2$, insufficient to destabilize the long-range ordered N\'eel phase. It is thus clear that a more thorough analysis is called for,
this being our main goal in the present work.

We employ three state-of-the-art methods to derive effective spin models for the Hubbard model on the honeycomb lattice at half-filling, in order to identify dominant
effective interactions as a function of $t/U$. Interestingly, among the vast number of effective spin couplings shooting up in the vicinity of the quantum phase
transition to the semi-metal, we find that the largest correction to the NN Heisenberg interaction are six-spin interactions located on single hexagons; the
frustrating next-NN exchange $J_2$ is considerably smaller. We assess the accuracy of the effective low-energy theory, for $t/U$ yielding the QSL,
by performing exact diagonalizations (EDs) on small clusters.

We organize the paper as follows. In Sec.~\ref{Model} we write down the Hubbard Hamiltonian and a generic effective spin model, accordingly fixing the
notation employed throughout the paper. The methods used in deriving the low-energy spin model, namely {\em perturbative} and
{\em graph-based} continuous unitary transformations (respectively, pCUTs and gCUTs), as well as the contractor renormalization (CORE) group, are
described in Sec.~\ref{Methods}. After comparing the outcomes from both approaches in Sec.~\ref{SpinModel}, we present results from EDs in Sec.~\ref{ed}.
Finally, we summarize our results in Sec.~\ref{Conclusions}.

\section{Model}
\label{Model}

We consider the single-band Hubbard model studied in Ref.~\cite{Meng10}, defined on a honeycomb geometry and at half-filling, that reads
\begin{equation}
\mathcal{H} = H_U+H_t
            = U\sum_{i}n_{i\uparrow}n_{i\downarrow} -t\sum_{\langle i,j\rangle,\sigma}(c_{i\sigma}^{\dagger}c_{j\sigma}+\textrm{h.c.})~.
\label{Hubbard_model}
\end{equation}
$n_{i\sigma}=c_{i\sigma}^{\dagger}c_{i\sigma}$ is the occupation number operator for fermions with spin $\sigma$ at the site $i$ of the honeycomb
lattice, and $t$ is the amplitude for hoppings taking place between NN sites, ${\langle i,j\rangle}$, on this lattice. Quantum Monte Carlo (QMC) simulations
\cite{Meng10} show that a QSL phase is stabilized for moderately strong couplings, $0.233 \lesssim t/U \lesssim 0.286$ ($4.3 \gtrsim U/t \gtrsim 3.9$), and it is our main purpose here to
compute effective spin interactions for Eq.~(\ref{Hubbard_model}) in this range.

Due to the $SU(2)$ symmetry of the Hubbard Hamiltonian, a generic effective model for its Mott insulating phase can be expressed in terms of products
of spin-half operators $( \vec{S}_{i} \cdot \vec{S}_{j} )$ as
\begin{eqnarray}
 H_{\rm spin} = E_0 &+& \sum_{i,j} J_{ij} \left( \vec{S}_{i} \cdot \vec{S}_{j} \right) + \sum_{i,j,k,l} K_{ijkl} \left( \vec{S}_{i} \cdot \vec{S}_{j} \right)
                            \left( \vec{S}_{k} \cdot \vec{S}_{l} \right) \\ \nonumber
                            &+& \sum_{i,j,k,l,m,n} L_{ijklmn} \left( \vec{S}_{i} \cdot \vec{S}_{j} \right)  \left( \vec{S}_{k} \cdot \vec{S}_{l} \right) \left( \vec{S}_{m}
                            \cdot \vec{S}_{n} \right)  + \ldots~,
\end{eqnarray}  
where $E_0$ denotes a constant energy shift. $J_{ij}$, $K_{ijkl}$ and $L_{ijklmn}$ respectively denote coupling constants for the various two-, four-, and
six-spin exchanges and the $\ldots$ refer to analogue expressions involving more than six spins. Unfortunately, however, the situation is complicated 
by the fact that multi-spin exchanges involving eight or more spins form an over-complete set of operators \cite{klein1980}. As a consequence, no unique
 solution in terms of spin operators can be obtained. While the effective Hamiltonian remains well defined in terms of matrix elements in the spin basis, 
 it is only the algebraic representation of the effective Hamiltonian in terms of spin operators that is not unique.

In the next section we will describe different approaches to derive such effective low-energy models. Furthermore, we extract and  discuss the most important corrections to the nearest-neighbor Heisenberg exchange giving rise to a minimal magnetic model capturing the essential physics of the full Mott phase. This minimal model as well as a more elaborate effective spin model is then analysed afterwards.  

\section{Methods and Effective Models}
\label{Methods}

In this section, we discuss details concerning the numerical methods employed in obtaining the effective spin interactions for the Hubbard model, Eq.~(\ref{Hubbard_model}), at half-filling and in the regime of strong to intermediate couplings. We start by the so-called continuous unitary transformations which allow us to gain a global view on the effective spin model and its most important spin interactions. Afterwards, we apply the contractor renormalization technique to confirm the behaviour of the dominant effective spin couplings. We find that both methods essentially agree when considering the same minimal set of considered clusters in both calculations. The latter motivates the definition of a minimal magnetic model. 

\subsection{CUTs and full effective spin model}
\label{sec:gcut}

We have calculated the dependence of the magnetic exchange couplings on $t/U$ using perturbative continuous unitary transformations (pCUTs)~\cite{Stein97,
Knetter00,Knetter03_1} along the lines of Ref.~\cite{Yang10,Reischl04}. The pCUT provides the magnetic couplings as series expansions in $t/U$. Note that the spectrum of the
Hubbard model is symmetric at half filling under the exchange $t\leftrightarrow -t$ and therefore only even order contributions are present \cite{MacDonald88}.
We have determined all two-spin, four-spin and six-spin interactions up to order 14. In order 14 one has to calculate 345 topologically different graphs. 

The bare series do not converge in the spin-liquid region of the Mott phase \cite{Yang11}. This is different to the recently analysed case of the Hubbard model 
 on the triangular lattice \cite{Yang10,Yang11} where the intermediate non-magnetic phase is already realized for smaller ratios of the bandwidth $W$ to 
the interaction $U$. It is therefore mandatory to apply resummation schemes of the series which is very complicate to be performed reliably for the 
full set of exchange couplings.  We have therefore applied the recently developped graph-based continuous unitary transformations (gCUTs) \cite{Yang11}
 in order to derive non-perturbatively the effective spin model. In the following we use the gCUT results to study the physics of
the full low-energy spin model. The pCUT results are only used to analyze the most important terms in the effective model.  
 
The general properties of the gCUT are discussed in Ref.~\cite{Yang11}. Here we only mention the specific approach for the current problem. The basic idea is to use a CUT to map $\hat{H}$ unitarily to an effective Hamiltonian $\hat{H}_{\rm eff}$ which has the special property that the block without double occupancies representing the effective low-energy spin model is decoupled from the rest of the Hamiltonian. The convergence of the gCUT is not triggered by a small parameter but it relies on the fact that the correlation length of charge fluctuations is finite in the Mott insulator. 

In the gCUT one generates all topologically distinct connected graphs $\mathcal{G}_\nu$ of the lattice and one sorts them by their number of sites $n$. On each
graph $\mathcal{G}_\nu$ a CUT is implemented by setting up the finite number of flow equations \cite{Wegner94,Glazek93,Glazek94}. A continuous auxiliary
variable $l$ is introduced defining the $l$-dependent Hamiltonian $\hat{H}^{{\mathcal{G}}_\nu} (l)$:=$U^\dagger (l)\hat{H}^{{\mathcal{G}}_\nu}U(l)$. Then the flow
equation is given by 
\begin{equation}
 \partial_l \hat{H}^{{\mathcal{G}}_\nu}(l) = [\hat{\eta} (l), \hat{H}^{{\mathcal{G}}_\nu}(l)] \quad ,
\end{equation} 
where $\hat{\eta} (l)$:=$-U^\dagger(l)(\partial_l U(l))$ is the anti-Hermitian generator of the unitary transformation. At the end of the flow $l=\infty$ one obtains an
effective graph-dependent Hamiltonian $\hat{H}_{\rm eff}^{{\mathcal{G}}_\nu}$. 

We have used the generator introduced by Wegner \cite{Wegner94}
\begin{equation}
 \hat{\eta}^{\rm Wegner} (l) = \left[  \hat{H}^{\rm d} (l) ,  \hat{H}^{\rm nd} (l) \right] \quad ,
\end{equation}
where the diagonal part $\hat{H}^{\rm d}$ is given by all matrix elements between states with the same number of double occupancies and $\hat{H}^{\rm nd}$ denote
all remaining non-diagonal processes. We stop the flow for each graph once the residual off-diagonality (ROD) is below $10^{-9}$. Here the ROD is defined as the
square root of the sum of all non-diagonal elements which connect to the low-energy subspace. Finally, one obtains the effective spin model $\hat{H}_{\rm spin}$ in
the thermodynamic limit by substraction of subcluster contributions and by embedding back the graphs on the honeycomb lattice \cite{Yang11}.  

The numerical effort scales exponentially with the number of sites $n$ of a given graph. Here we have treated all graphs up to 7 sites. This amounts to solving up to one million differential equations for the most demanding symmetry sector of a single graph for each value of $t/U$. Physically, one captures all charge fluctuations on this length scale. Let us remark that it is of course possible to only include a restricted set of graphs in the gCUT calculation. This will be used below for a direct comparison with CORE. Finally, one obtains for a given $t/U$ the effective spin model non-perturbatively.

\begin{figure}
\begin{center}
\includegraphics*[width=0.75\columnwidth]{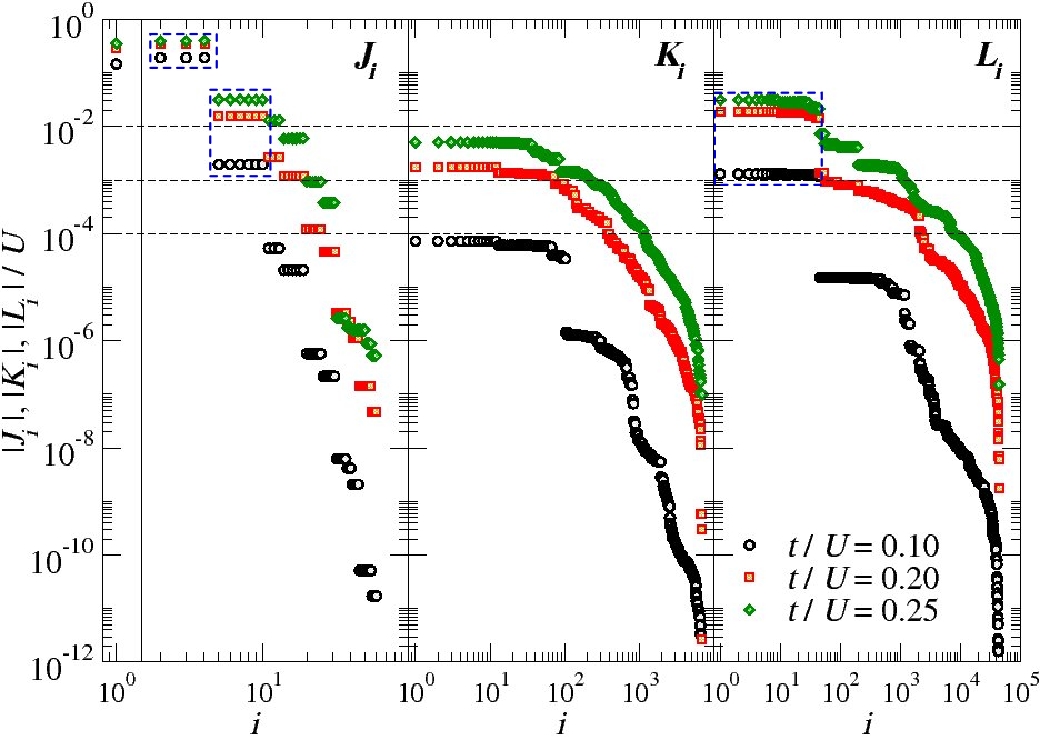}
\end{center}
\caption{(Color online) Coefficients of all gCUT(7) terms in the effective Hamiltonian for different $U/t$ in units $1/U$. Terms with different number of spin operators 0, 2, 4, or 6 are separated by vertical solid lines. Each block containing the same number of spin operators is sorted by the size of the coefficient. Here each spatially symmetric coupling is also shown. Dashed blue boxes highlight the most important couplings. Most left box refers to the three equivalent couplings $J_1$, the middle box corresponds to the six equivalent couplings $J_2$, and the most right box includes the six-spin interactions $L_1,L_2,\cdots,L_5$ on a hexagon.} 
\label{fig:couplings_gCUT_overview}
\end{figure}

A view on the full effective spin model obtained by gCUTs for different values of $t/U$ is shown in Fig.~\ref{fig:couplings_gCUT_overview}. Let us remark that for gCUT(7)
 the full low-energy model can be uniquely expressed in terms of spin operators, because multi-spin interactions involving more than six sites are not yet allowed for this cluster size. 

There are several implications which can be directly read off from Fig.~\ref{fig:couplings_gCUT_overview}. First, only at rather small values of $t/U \lesssim 0.1$ a clear hierarchy in the amplitudes of the effective spin operators can be seen. This corresponds to the purely perturbative regime where the importance of terms is fully given by the perturbative order where a spin coupling appears for the first time. In this regime the frustrated next-nearest neighbor Heisenberg exchange $J_2$ (order 4 perturbation theory) is the leading correction to the dominant Heisenberg interaction $J_1$ (order 2 perturbation theory). 

Second, the situation changes drastically in the intermediate coupling regime where the spin-liquid phase is realized. Here, no clear hierarchy in the amplitudes of the effective spin couplings can be detected. One is rather confronted with a proliferating number of terms which is likely a consequence of the fact that the metal-insulator transition is second order. Consequently, the charge gap closes continuously and one expects an increasing number of effective spin couplings on increasing length scales when approaching the quantum critical point. Let us note that this is different for the Hubbard model on the triangular lattice \cite{Yang10}. In this case the metal-insulator transition is first order and one has no diverging length scale 
upon approaching the Mott transition from the insulating side.

Third, and despite the complexity of the full effective spin model, one still expects that most of the properties of the Mott phase should be contained in a minimal magnetic model where one only considers the most important spin couplings having the largest amplitudes. Some evidences for this strategy are given in the next section by analysing such a minimal magnetic model. Interestingly, we find that the perturbative hierarchy is not valid anymore  for intermediate couplings: the most important correction to $J_1$ is not $J_2$ but rather six-spin interactions located on hexagons. These six-spin interactions arise in order six perturbation theory and they originate dominatly from ring exchange processes around the shortest loop of even length on the honeycomb lattice corresponding to hexagons. Let us remark that this is very similar to the Hubbard model on the triangular lattice \cite{Yang10}. Here the most elementary loop of even length is a four-site plaquette. Consequently, four-spin interactions on the elementary plaquette are the dominant correction to the Heisenberg model. 

\begin{figure}
\begin{center}
\includegraphics*[width=0.3\columnwidth,angle=270]{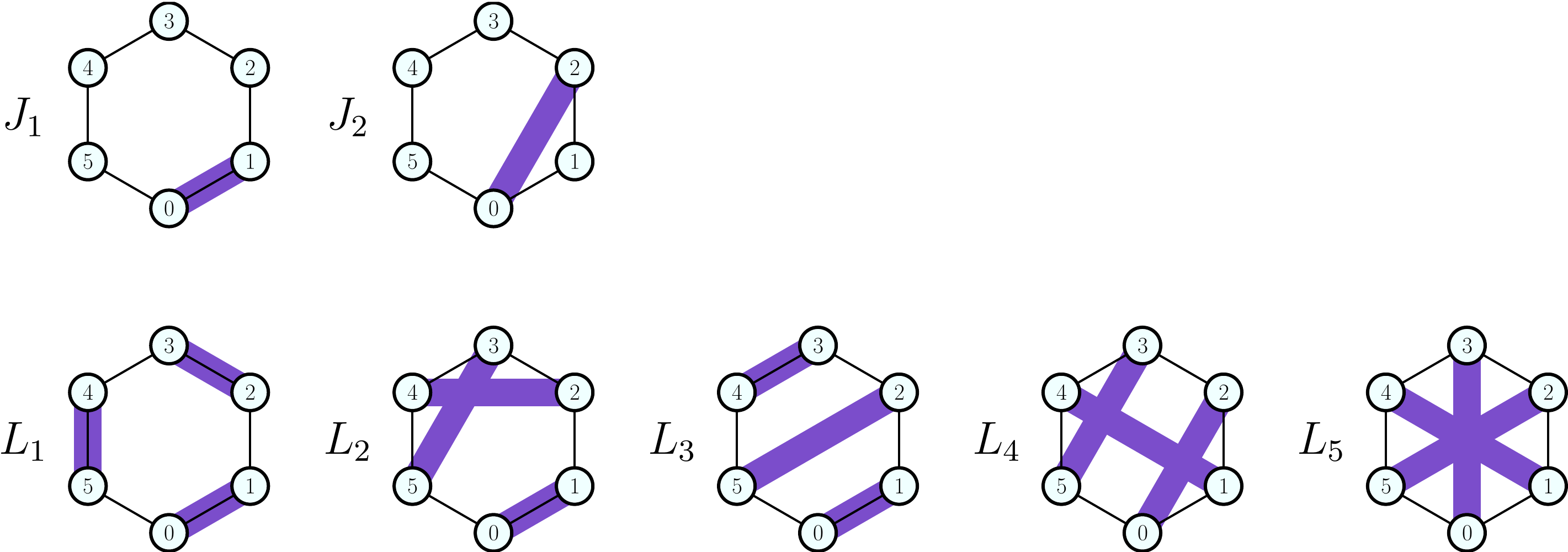}
\end{center}
\caption{(Color online) Dominant effective spin interactions for the Hubbard model on the honeycomb lattice at half-filling [Eq.~(\ref{Hubbard_model})]:
                NN and next-NN two-spin couplings ($J_1$ and $J_2$, respectively), and six-spin  exchange terms with couplings
                $L_1,L_2,\cdots,L_5$.} 
\label{fig:couplings}
\end{figure}

\begin{figure}
\begin{center}
 \includegraphics*[width=0.4\columnwidth]{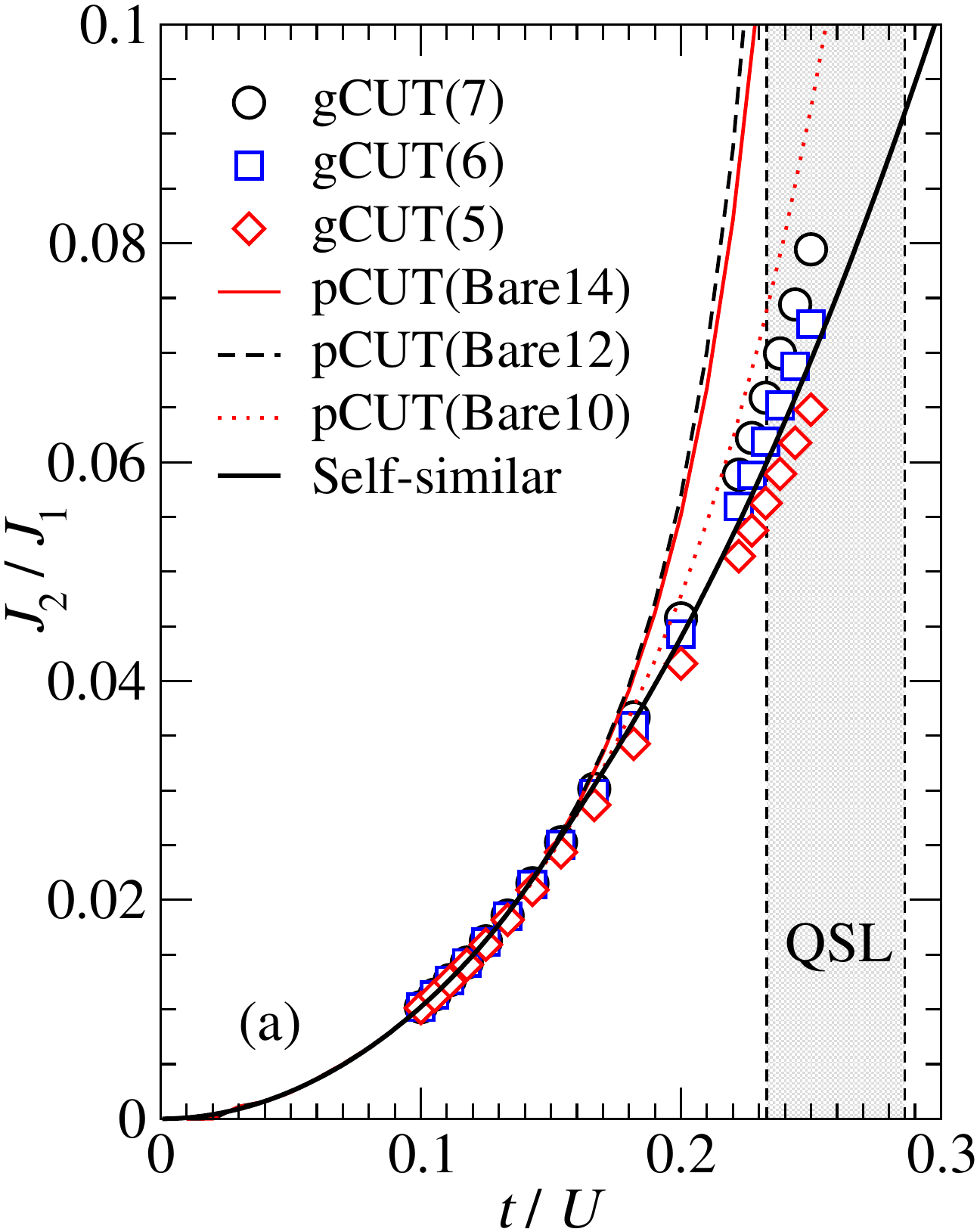}
 \hspace{0.8cm}
 \includegraphics*[width=0.4\columnwidth]{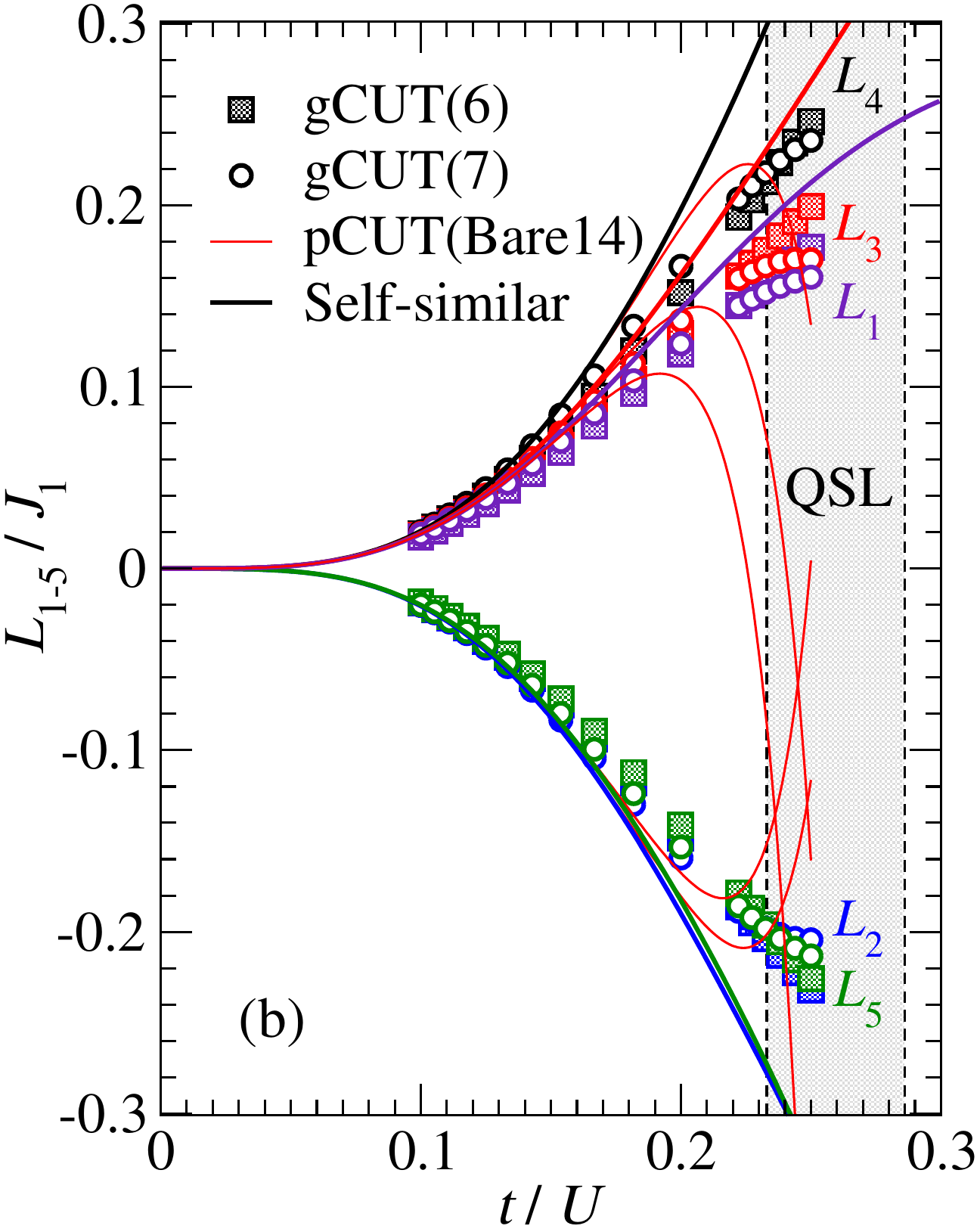} 
\end{center}
\caption{(Color online) The relative gCUT and pCUT amplitudes $J_2/J_1$ (left) and $L_i/J_1$ (right) as a function of $t/U$. Thin solid lines refer to the bare pCUT series. Black solid lines correspond to self-similar extrapolants as discussed in Ref.~\cite{Yang11}. Symbols denote results obtained by gCUTs.} 
\label{fig:couplings_gCUT_sixspin}
\end{figure}

In the remainder of this paper we are aiming at a minimal magnetic model which contains only the most important spin interactions in the regime of strong to intermediate coupling. In this parameter regime, the minimal model should display the low-energy properties of the Hubbard model on the honeycomb lattice. Additionally, we expect that going beyond the minimal model by considering the full graph decomposition used by gCUT (or alternatively by CORE) yields successfully even better results. The most important effective exchanges are depicted in Fig.~\ref{fig:couplings}: NN ($J_1$) and next-NN ($J_2$) two-spin interactions appearing in the $J_1$-$J_2$ model, and the six-spin terms with couplings $L_1,L_2,\cdots,L_5$. The corresponding gCUT and pCUT amplitudes for these couplings are displayed in Fig.~\ref{fig:couplings_gCUT_sixspin}.

One clearly sees that the bare series of the six-spin interactions are not converged in the interesting regime of intermediate $t/U$ confirming the conclusion obtained in Ref.~\cite{Yang11}. In contrast, the results obtained by gCUT and self-similar extrapolation of the pCUT series are in fair agreement for all displayed couplings. 

In the following we want to further strengthen these findings by using CORE as an alternative tool to derive effective low-energy models. 

\subsection{Contractor renormalization and minimal model}
\label{sec:core}

It is also possible to derive non-perturbative effective theories by relying on the CORE technique~\cite{CORErefs}. CORE is a real-space renormalization procedure,
applicable to generic lattice models, where effective models are derived by truncating {\em local} degrees-of-freedom. Such construction requires the exact computation
of low-lying eigenstates on finite connected graphs, similarly to what happens in the previously discussed gCUT procedure; implementation details are discussed in the
literature \cite{CORErefs,Capponi2004}. The resulting effective Hamiltonian is given as a cluster expansion 
\begin{equation}
{\cal H}_{\mathrm{eff}}= \sum_{g} h_g^c~,
\end{equation}
where the sum takes place over a set of graphs $g$, and $h_g^c$ corresponds to the \emph{connected} term that is obtained by substracting contributions from
embedded sub-clusters \cite{CORErefs,Capponi2004}. Such expansion must naturally be truncated at a certain maximum {\em range} \cite{Siu2007}, but its
accuracy can in principle be controlled by analyzing the convergence of terms in the effective model with increasing range.

\begin{figure}
\begin{center}
\includegraphics*[width=0.175\columnwidth,angle=270]{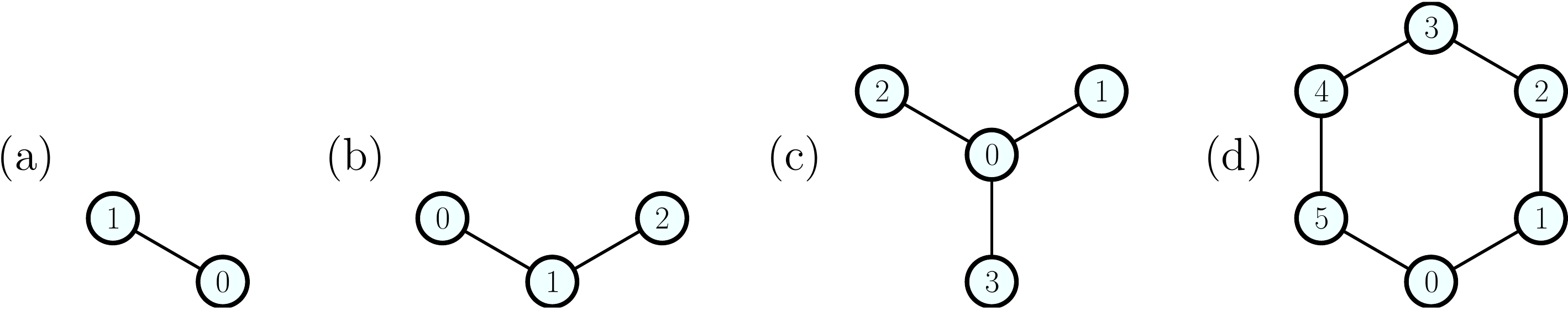}
\end{center}
\caption{Clusters considered in the CORE expansion, ordered according to the maximum range for effective interactions \cite{Siu2007} and comprising: (a) two, (b)
               three, (c) four and (d) six sites.} 
\label{fig:CORE_clusters}
\end{figure}

Since it is our goal here to obtain effective {\em spin} interactions for the Hubbard model [Eq.~(\ref{Hubbard_model})] at half-filling, we select single sites as {\em elementary
blocks} in the CORE expansion \cite{CORErefs,Capponi2004} and retain only singly-occupied states on all sites. 

The clusters used in the CORE calculation are shown in Fig.~\ref{fig:CORE_clusters}. Here two setups are considered: i) The first choice of clusters does not contain the four-site star graph displayed in Fig.~\ref{fig:CORE_clusters}(c). It therefore corresponds to the leading term of a full graph expansion in terms of hexagons. ii) The second choice contains additionally the four-site star cluster [Fig.~\ref{fig:CORE_clusters}(c)]. In the hexagon expansion mentioned before such a graph would be included only on the level of three-hexagon clusters. 
Later we see that gCUT (restricted to exactly the same graphs) yields almost identical results for both choices of clusters. The second choice can then be regarded as an intermediate step between the minimal one-hexagon calculation and the full gCUT(7) calculation which includes all graphs shown in Fig.~\ref{fig:CORE_clusters} as a subset.

\begin{figure}
\begin{center}
\includegraphics*[width=0.6\columnwidth,angle=270]{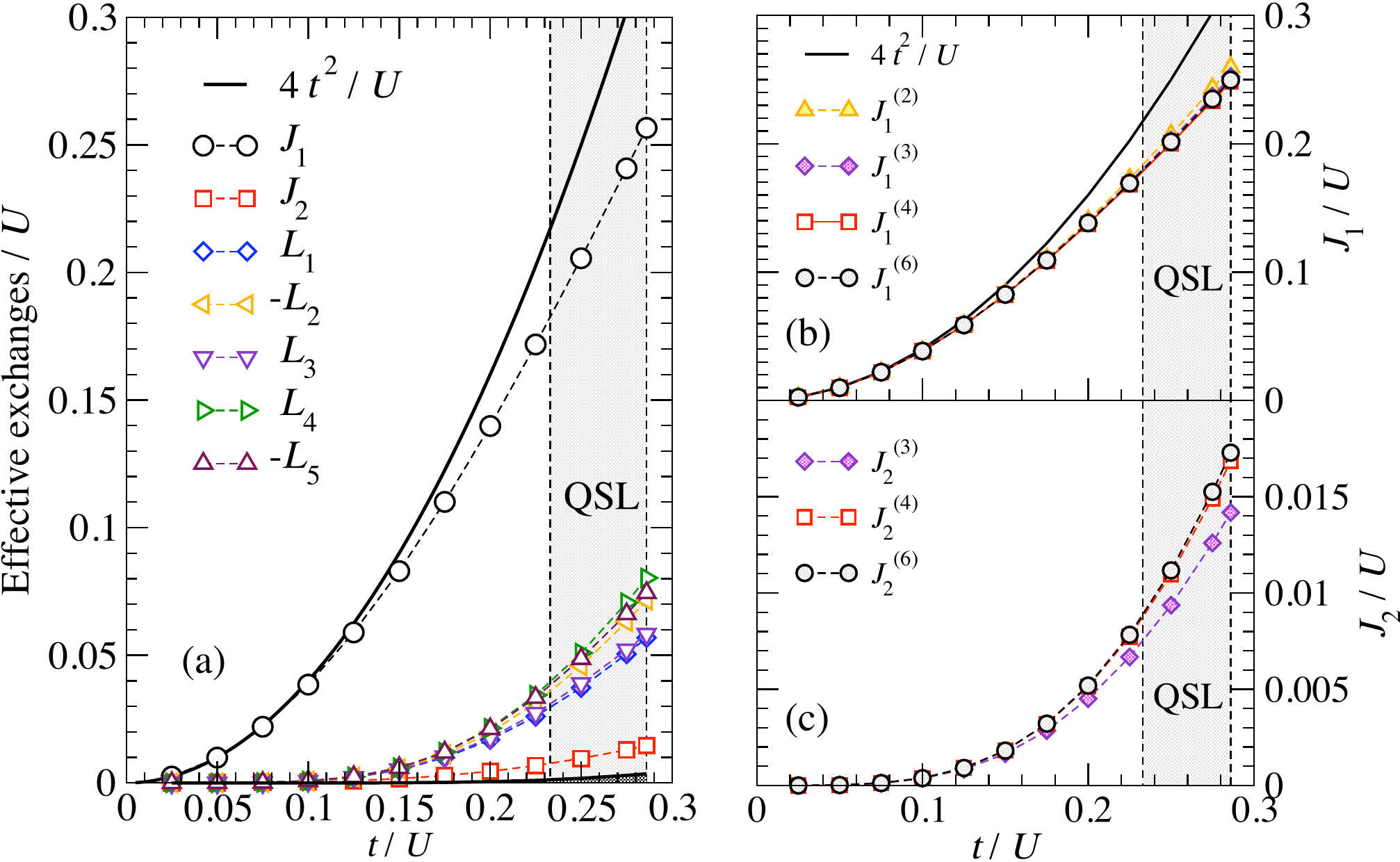}
\end{center}
\caption{(Color online) (a) {\em Bare} amplitudes for of all terms in the effective Hamiltonian for Eq.~(\ref{Hubbard_model}) as a function of $t/U$, as obtained from the
                CORE procedure applied to the six-site cluster depicted in Fig.~\ref{fig:CORE_clusters}(d). For clarity sake, only dominant couplings are shown, following
                the notation from Fig.~\ref{fig:couplings}; all remaining terms have very small amplitudes and lie in the filled region at the bottom-left corner.
                (b-c) Results for the two-spin exchanges $J_{1}$ and $J_{2}$, as a function of $t/U$, obtained from indicated clusters (we label the results according to
                the number of sites in a given cluster; see Fig.~\ref{fig:CORE_clusters}). For each cluster, contributions from embedded clusters comprising lesser sites
                are subtracted, following the standard CORE recipe \cite{CORErefs,Capponi2004}. In both panels, the shaded region corresponds to the QSL phase
                \cite{Meng10}.
} 
\label{fig:couplings_CORE}
\end{figure}

We start by discussing the first choice. The lowest-order contribution to the NN spin-exchange $J_1$ is simply computed by solving the Hubbard model on a two-site cluster [Fig.~\ref{fig:CORE_clusters}(a)]: in this case only eigen-energies are required, since $J_1$ {\em exactly} corresponds to the singlet-triplet gap:
\begin{equation}
J^{(2)}_1=\frac{U}{2} \left( \sqrt{1+\frac{16 t^2}{U^2}}-1 \right)~.
\end{equation}
Although trivially obtained, this result is non-perturbative, reducing to the well-known limit $J_1 \simeq 4t^2/U$ only when $t/U \ll 1$ [see Fig.~\ref{fig:couplings_CORE}(b)]. Let us mention that the same non-perturbative result is obtained when calculating $J^{(2)}_1$ with gCUT(2).

Longer-range couplings, along with corrections to shorter-ranged ones, are obtainable from the analysis of the larger clusters depicted in Fig.~\ref{fig:CORE_clusters}(b-d). In particular, in Fig.~\ref{fig:couplings_CORE}(a) we plot {\em bare} effective couplings, obtained by considering the six-site cluster in Fig.~\ref{fig:CORE_clusters}(d), as a function of $t/U$. Here the notion 'bare couplings' corresponds to the effective exchange couplings of a single cluster without considering substractions and embeddings. We first focus on the next-NN coupling $J_{2}$ and notice that, particularly in the range $0.233 \lesssim t/U \lesssim 0.286$ (($4.3 \gtrsim U/t \gtrsim 3.9$)) where a QSL appears
\cite{Meng10}, $J_{2} \lesssim 0.1 J_{1}$ and therefore this frustrating interaction is not large enough \cite{Albuquerque11,Reuther11,Oitmaa11,Mosadeq11,Mezzacapo12}
to account for the absence of long-range N{\' e}el order before a semi-metal phase is stabilized \cite{Yang11}. However, we also notice that, remarkably, much larger magnitudes are
associated to the six-spin terms with couplings $L_1,L_2,\cdots,L_5$ depicted in Fig.~\ref{fig:couplings} (we also briefly remark that the signs for their amplitudes are in
agreement with those obtained from a decomposition of the cyclic 6-spin permutation, cf.~Ref.~\cite{Godfrin1988}). Furthermore, we observe that the amplitudes for longer-range two-spin, as well as four-spin, terms are found to be
negligible in comparison to $J_1$ and $L_1,L_2,\cdots,L_5$ \cite{remark_cyclic}. One is thus inclined to conclude that such six-spin interactions are the main ingredients in stabilizing a QSL in the phase diagram for Eq.~(\ref{Hubbard_model}), a conjecture we aim at testing in Sec.~\ref{ed}. 

Let us mention that such six-spin interactions on local hexagons are generated in order six perturbation theory. In this order the couplings $L_1$-$L_5$ have the same absolute prefactor with an alternating sign which exactly corresponds to the six-spin interactions contained in the six-spin ring exchange permutation operator. But we stress that the latter operator also contains two-spin and four-spin interactions of the same magnitude which one has to contrast with our finding, both perturbatively and non-perturbatively, of suppressed two- and four-spin interactions. The full non-perturbative contribution of a single hexagon is therefore dominated by the six-interactions having the largest exchange couplings. Finally, we note that the nonperturbatively obtained couplings $L_1$-$L_5$ have different prefactors as can be seen in Fig.~\ref{fig:couplings_CORE}. 

We proceed by analyzing the dependence of the effective couplings on the maximum range of the clusters. In order to do so, for each cluster depicted in Fig.~\ref{fig:CORE_clusters}
we subtract {\em connected} contributions from shorter-ranged, embedded, sub-clusters, according to the standard CORE recipe \cite{CORErefs,Capponi2004}. We plot
results for the two-spin exchanges with couplings $J_{1}$ and $J_{2}$ as a function of $t/U$ respectively in Figs.~\ref{fig:couplings_CORE}(b) and (c), and find a seemingly
fast convergence with increasing cluster range up to the six-site cluster depicted in Fig.~\ref{fig:CORE_clusters}(d). Furthermore, the data in Figs.~\ref{fig:couplings_CORE}(b) and (c) is in good agreement with gCUT results, as we discuss below in Sec.~\ref{SpinModel}. 

Finally, we comment on the fact that a few clusters larger than the ones in Fig.~\ref{fig:CORE_clusters}, such as the one comprising ten sites and forming an edge-sharing double hexagon, are amenable to numerical diagonalization, in principle implying that our CORE expansion could be extended to even longer ranges than considered here. To this end a full graph decomposition must be implemented for the CORE approach which will be left as a task for future research. Here we would rather like to focus on the most minimal setup to describe the low-energy physics of the Hubbard model.

\subsection{Minimal effective model}
\label{SpinModel}

\begin{figure}
\begin{center}
 \includegraphics*[width=0.75\columnwidth]{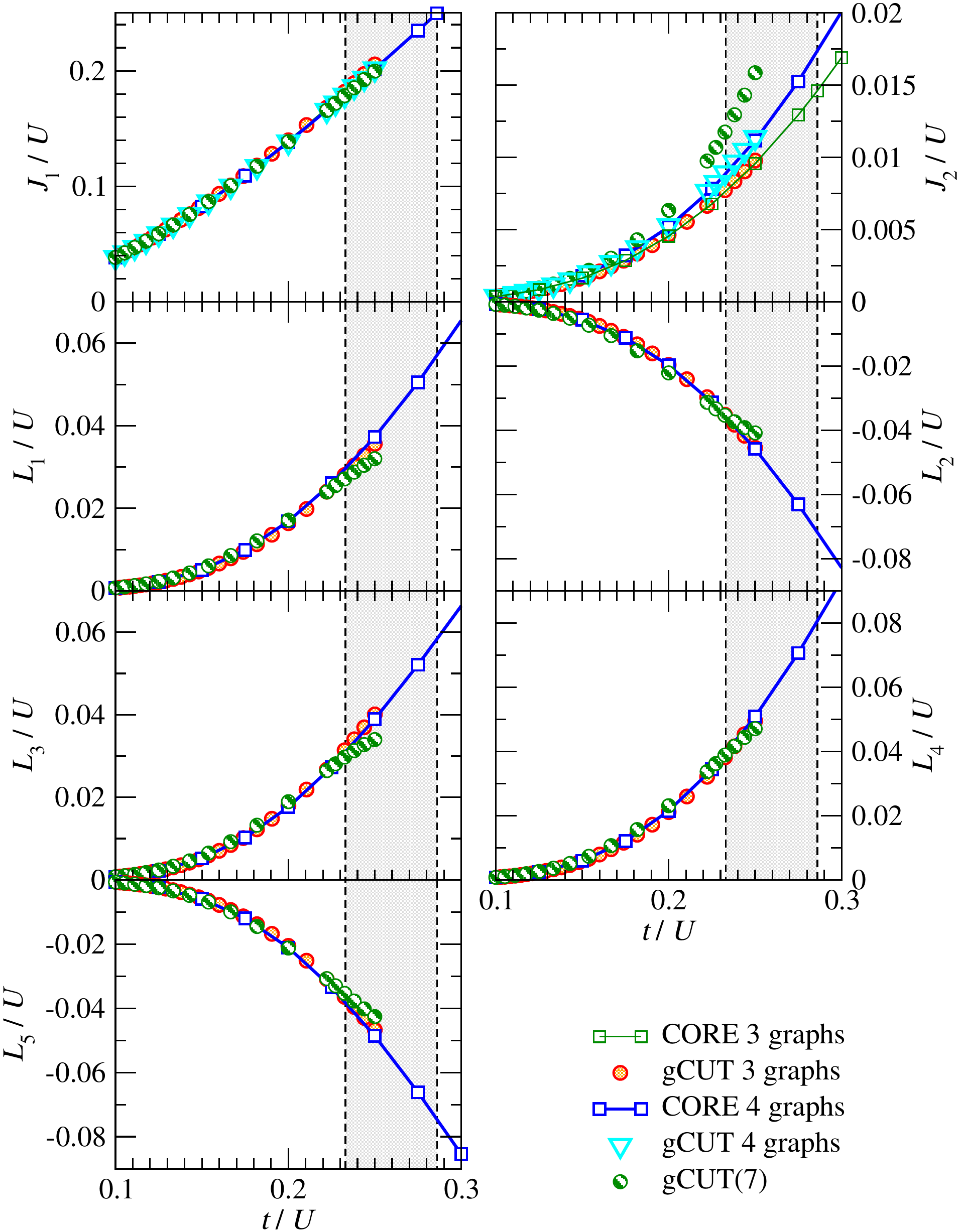}
\end{center}
\caption{(Color online) Comparison between effective couplings, obtained from gCUT (Sec.~\ref{sec:gcut}) and CORE (Sec.~\ref{sec:core}), for the dominant spin-exchanges
                                          depicted in Fig.~\ref{fig:couplings}. The range of values for $t/U$ leading to the QSL phase, according to Ref.~\cite{Meng10}, is highlighted.} 
\label{fig:comparison}
\end{figure}

We compare now the effective models obtained from gCUT (Sec.~\ref{sec:gcut}) and CORE (Sec.~\ref{sec:core}). We first remark that both approaches yield qualitatively similar effective Hamiltonians, that display as dominant terms the two- and six-spin exchanges depicted in Fig.~\ref{fig:couplings} (see Figs.~\ref{fig:couplings_gCUT_sixspin}
and \ref{fig:couplings_CORE}). We therefore focus on such terms, that may be regarded as defining a {\em minimal effective model}, and plot their amplitudes, as a function of $t/U$ in Fig.~\ref{fig:comparison}. Let us mention that we do not show the constant term of the effective models which also displays a $t/U$ dependence and which one therefore has to keep in the exact diagonalization when comparing to the Hubbard model. Here three different sets of graphs are considered: i) the leading graphs of a hexagon expansion not including the four-site star graph (CORE and gCUT), ii) the leading graphs of a hexagon expansion including the four-site star graph (CORE and gCUT), and iii) all graphs comprising up to seven sites (gCUT).

We notice that results from both methods are in good quantitative agreement up to couplings stabilizing the QSL phase \cite{Meng10} and beyond the perturbative regime \cite{Yang11} for all three choices. Especially, CORE and gCUT yield almost identical results for the restricted graph sets i) and ii). This is likely a consequence of the fact that the structure of the low-energy subspace is rather constraint due to the high symmetry of the considered clusters (the four-site graph as well as the six-site graph has a rotational symmetry) plus the number of different spin couplings defined on these graphs is rather small.

The most noticeable effect when going from the minimal choice to the full graph decomposition including all graphs up to seven sites is an increasing value for the next-NN two-spin exchange $J_2$. Nevertheless, we also remark all three sets agree in that, unlike what is assumed in Refs.~\cite{Wang2010,Cabra11}, the effective values for the next-NN two-spin exchange $J_2$ are not large enough to destroy N{\' e}el order \cite{Albuquerque11,Reuther11,Oitmaa11,Mosadeq11,Mezzacapo12}, suggesting that the QSL phase may instead be accounted for by the six-spin ring exchanges with amplitudes $L_1,L_2,\cdots L_5$ in Fig.~\ref{fig:couplings}. Furthermore, we show in the next section that the minimal effective spin model having the lowest $J_2$ give the best agreement with the QMC results at intermediate couplings.

\section{Exact diagonalizations}
\label{ed}

We turn now our attention to the characterization of the effective
spin model derived in the previous sections, either with gCUT method
or CORE algorithm. Our goal is to show that a reasonably simple and
compact effective spin model can describe rather well the energetics
of the Hubbard model. In order to do so, we will make a systematic
comparison of the low-energy properties, and of the local quantities
such as double occupancy or bond kinetic energy, that we
compute using the exact diagonalization (ED) technique. More
precisely, we have used a standard Lanczos algorithm to obtain the
ground-state energy for various clusters at fixed total $S_z$, and we
have made use of all space symmetries (translations and point-group,
when available).

\subsection{Study of the minimal effective model}

We start by discussing the properties of the minimal effective spin model 
obtained for the minimal set of graphs not including the four-site star graph.
We stress again that CORE and gCUT give essentially identical low-energy spin
 models for this choice of graphs. In the following we use the effective model 
 obtained via the CORE algorithm (Sec.~\ref{sec:core}). 

\begin{figure}
\begin{center}
 \includegraphics*[width=0.525\columnwidth,angle=270]{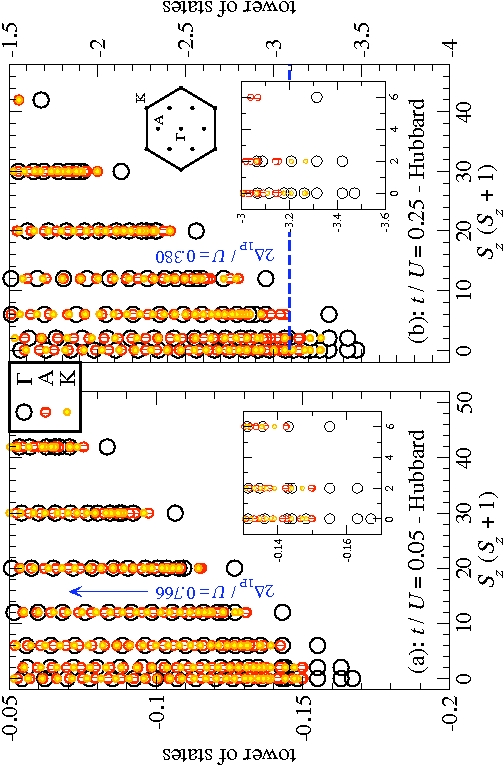}
 
  \vspace{0.5cm}

\includegraphics*[width=0.525\columnwidth,angle=270]{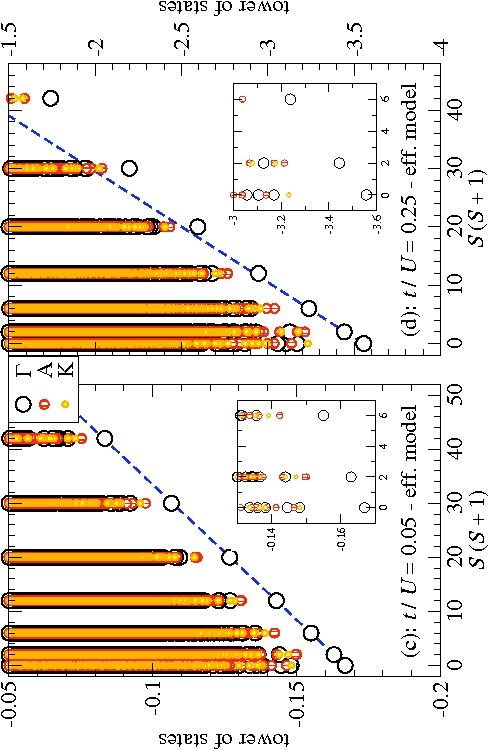}
\end{center}
\caption{(Color online) (a-b) Low-energy spectrum of the Hubbard model vs $S_z(S_z+1)$ obtained for $t/U=0.05$ (a) and $t/U=0.25$ (b) on an $N=18$ honeycomb cluster. Values for twice the one-particle gap, $2\Delta_{1P}$, are given [in (b), such value is indicated by the horizontal dashed line]. (c-d) Low-energy spectrum for the CORE effective model (see main text) vs $S(S+1)$, obtained on the same $18$-site cluster, again for $t/U=0.05$ (c) and $t/U=0.25$ (d). In all panels, different symbols correspond to different points in the Brillouin zone: $\Gamma$ point at the center, 6-fold degenerate $A$ point and 2-fold degenerate $K$ point at the corners [see inset in (b)]. Lowest energy levels in all cases are zoomed in in the insets.}
\label{fig:TowerN18}
\end{figure}

We start by comparing the low-energy spectra for the Hubbard
[Eq.~(\ref{Hubbard_model})] and the effective (Sec.~\ref{Methods})
models on an $N=18$ cluster for two values of $t/U$~: $t/U=0.05$ (deep
into the N\'eel phase) and $t/U=0.25$ (QSL phase, according to
Ref.~\cite{Meng10}). In Fig.~\ref{fig:TowerN18}(a-b), we plot
so-called Anderson tower of states for the Hubbard model. As we only
make use of $S_z$ symmetry (instead of the total spin), we plot the
energy states {\em versus} $S_z(S_z+1)$, so that degenerate states
with identical energies correspond to spin multiplets. We have also
computed the one-particle gap, $\Delta_{1P}=(E_0(N/2+1)+E_0(N/2-1)-
2E_0(N/2))/2$, where $E_0(N_f)$ is the ground-state energy with $N_f$
fermions. For $t/U=0.05$ [Fig.~\ref{fig:TowerN18}(a)], the so-called Anderson tower of
states is typical of a collinear N\'eel state~\cite{Misguich2007} 
 with a lowest excitation increasing as $S(S+1)$ (with a
slope that should decrease as $1/N$), followed by spin-wave
excitations (with energy scale $1/\sqrt{N}$). Moreover, on the scale
of the figure, all states correspond to spin excitations since
$\Delta_{1P}$ is quite large \cite{gap_remark}. The same behaviour is reproduced
qualitatively and quantitatively by the effective spin model, as
expected and shown in Fig.~\ref{fig:TowerN18}(c). Note that in the case of the effective spin model, we have been able to label each eigenstate with its total spin value $S$ so that we plot levels as a function of $S(S+1)$. 

When increasing $t/U$ to $0.25$, a distinct picture emerges
[Fig.~\ref{fig:TowerN18}(b)] . While the first excitations still
correspond to $S=1$ and $S=2$ states at the $\Gamma$ point, there is
no longer a straight $S(S+1)$ line of excitations and, even more
importantly, the gap $\Delta_{1P}$ to {\em charge} excitations is
substantially lowered, beyond which a description solely based on spin
degrees-of-freedom is no longer valid. However, according to QMC
results~\cite{Meng10}, for this coupling the one-particle gap should
remain finite in the thermodynamic limit, so that our effective spin
model could still describe the physics at the lowest energies: indeed,
its low-energy spectrum, plotted in Fig.~\ref{fig:TowerN18}(d),
reproduces the lowest spin excitations quite accurately. Furthermore, 
while the quantum numbers seen in the tower-of-states are consistent with standard collinear
magnetic ordering, the lowest spin excitations do not display a clear $S(S+1)$ behaviour, which might be an indication that there is no magnetic ordering.

\begin{figure}
\begin{center}
 \includegraphics*[width=0.8\columnwidth]{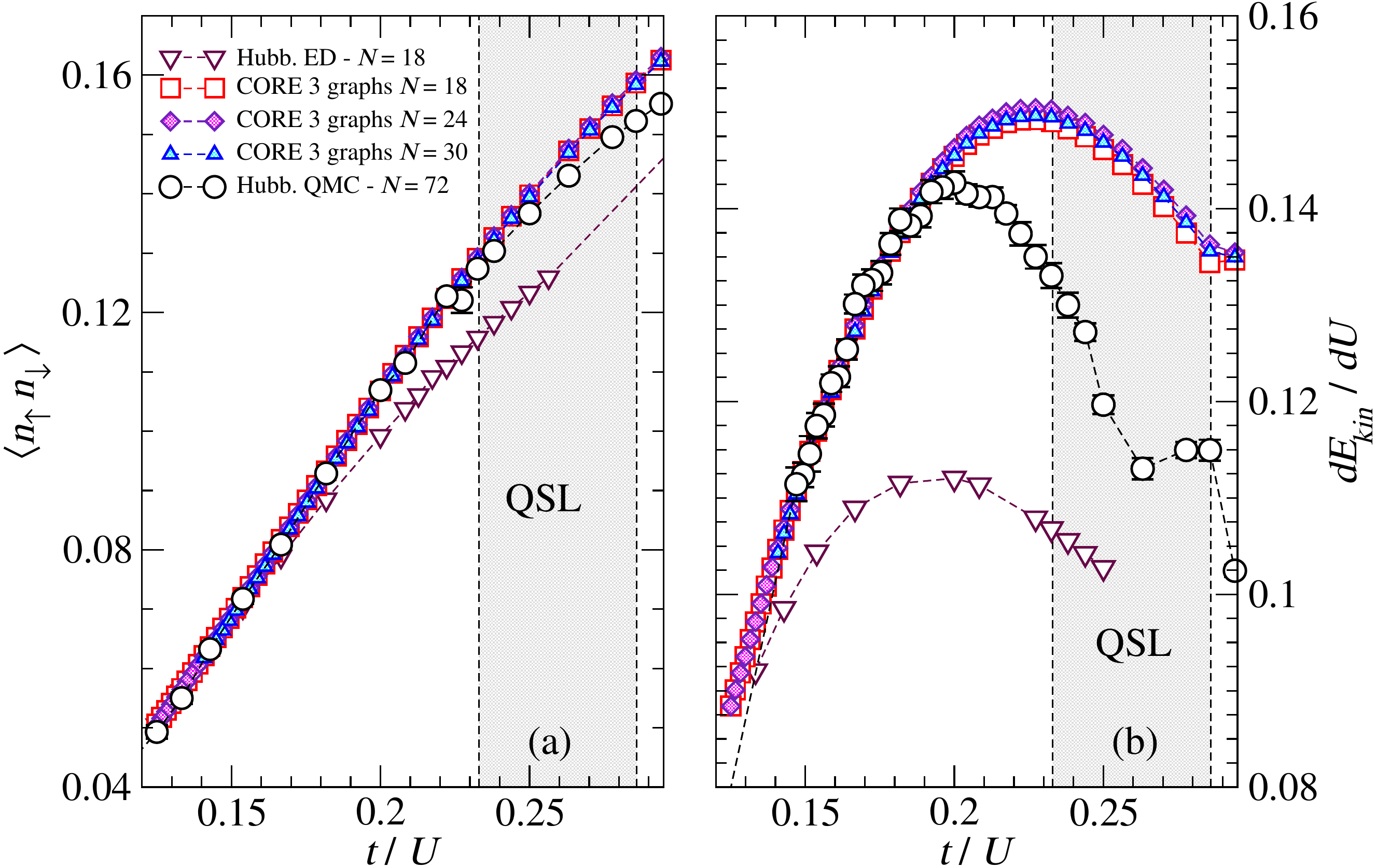}
\end{center}
\caption{(Color online) (a) Double occupancy per site obtained obtained with the minimal CORE effective model (3 graphs) on finite clusters with  $N=18$ to $N=30$. (b) Derivative of the kinetic energy, as a function of $t/U$, obtained from the ground-state energy of the effective model [Eq.~(\ref{Ekin})] on the same clusters.  
In both cases, comparison is made to ED ($N=18$) and QMC ($N=72$)~\cite{Thank_Meng} data of the Hubbard model.} 
\label{fig:kin_vs_U}
\end{figure}

It would also be desirable to compute correlators for the ground-state 
of the effective model, but this is quite involved since one would also
need to renormalize the operators when deriving an effective
Hamiltonian~\cite{CORErefs,Capponi2004,Yang10,Yang11}. Nevertheless, some
observables may straightforwardly be obtained from the ground-state
energy per site, $e_0$, by relying on the Feynman-Hellmann
theorem. For instance, the electronic double occupancy per site can be
computed from ED of the effective model involving only spin
degrees-of-freedom, as it has been done in Ref.~\cite{Yang10} using~:
\begin{equation}
\langle n_{\uparrow} n_\downarrow \rangle = \frac{d e_0}{dU}
\end{equation}
Results are shown in Fig.~\ref{fig:kin_vs_U}(a) and compared both to
the exact result of the Hubbard model on $N=18$ cluster and to QMC
data~\cite{Thank_Meng}. Since this is a local quantity, its finite
size effects are rather small as expected. We observe a good
agreement not only for small $t/U$ (deep in the N\'eel phase), but
also in the interesting regime $t/U \simeq 0.25$ where QSL is expected
to emerge. As a side remark, let us emphasize that the discrepancy
between results on the Hubbard model and with the CORE effective one
on $N=18$ can be attributed to the role of short-length loops (of
length 6) that exist on this small cluster, and are not captured by
the effective model (see Ref.~\cite{Yang10,Albuquerque11} for a similar
discussion). In fact, effective model results agree quite well with the
data obtained on a much larger lattice with QMC.

In Ref.~\cite{Meng10}, the analysis of the behaviour of the kinetic energy density $E_{kin}=\langle
-t \sum_{\langle ij\rangle,\sigma} (c^\dagger_{i\sigma} c_{j\sigma}+h.c.)\rangle/N$ {\em versus} $U$ was also discussed in the context of the formation of local moments at strong $U/t$,
in contrast with the itinerant regime for small $U/t$, whereas the QSL is stabilized in between. Using Feynman-Hellmann theorem again, we can simply compute this quantity as
\begin{equation}
\frac{d E_{kin}}{dU} = - U \frac{d^2 e_0}{dU^2}~.
\label{Ekin}
\end{equation}
We have chosen a grid of $U/t$ going from 3 to 8 by step of $0.1$ and
computed the ground-state energy for various clusters. Using a
finite-difference approximation, we obtain the derivative of the
kinetic energy that we plot against $t/U$ in
Fig.~\ref{fig:kin_vs_U}(b). Our first comment is that we have some
agreement with the exact QMC data~\cite{Meng10} that show a maximum
around 0.14 for $U/t \simeq 5.0$, although this quantity is expected
to be quite sensitive to details since it is a second derivative of
the ground-state energy. Note that QMC data are also quite noisy in
this large $t/U$ region, and we refer to the original
publication~\cite{Meng10} for additional data~: the point is that a
maximum of order 0.15 is reached for $t/U \simeq 0.2$. Our second
remark is that we observe an anomaly around $t/U=0.3$ which could
signal that our CORE truncation is not safe beyond this value. This
gives us confidence that at least for $t/U\simeq 0.25$, our approach could make
sense, and this includes the spin liquid phase. 
Last, since we observe this change of curvature around $U/t=4.5$ in a spin-only model, it
seems to us that the virtual charge fluctuations giving rise to our effective low-energy model
 are sufficient to account for this phenomenon. Additionally, it might be not surprising that we 
underestimate this effect by our minimal magnetic model, because we expect that the large number of 
neglected additional spin operators as well as small renormalizations of the treated spin couplings could well
 lead to an enhanced itineracy of the ground state at intermediate $t/U$ values.

As a conclusion on this part, while we are clearly not able to reach system lengths where QSL behavior is expected to occur -- which would require several hundred sites -- our simple effective model is able to reproduce both the low-energy properties and the ground-state local properties. It suggests that these properties may not be linked to the presence of a QSL and are robust short-distance features. 

\subsection{Study beyond the minimal  model}

The results of the last subsections are very promising. CORE and gCUT give an almost identical minimal effective spin model which is obtained from the single hexagon graph and which compares well to the results obtained by QMC for the Hubbard model on the honeycomb lattice. 
  
In the following we want to go beyond this minimal choice of graphs and we would like to see whether the above findings are confirmed and strengthened. We therefore compare the above results for the minimal model to exact diagonalizations for the effective spin models obtained i) by including the four-site star graph to the minimal set and ii) by including all graphs up to seven sites. For case i) we again use results from the CORE algorithm, but we stress that the gCUT on the restricted set of graphs gives basically the same low-energy model. For the second case the results of gCUT(7) are used. 

\begin{figure}
\begin{center}
 \includegraphics*[width=0.8\columnwidth]{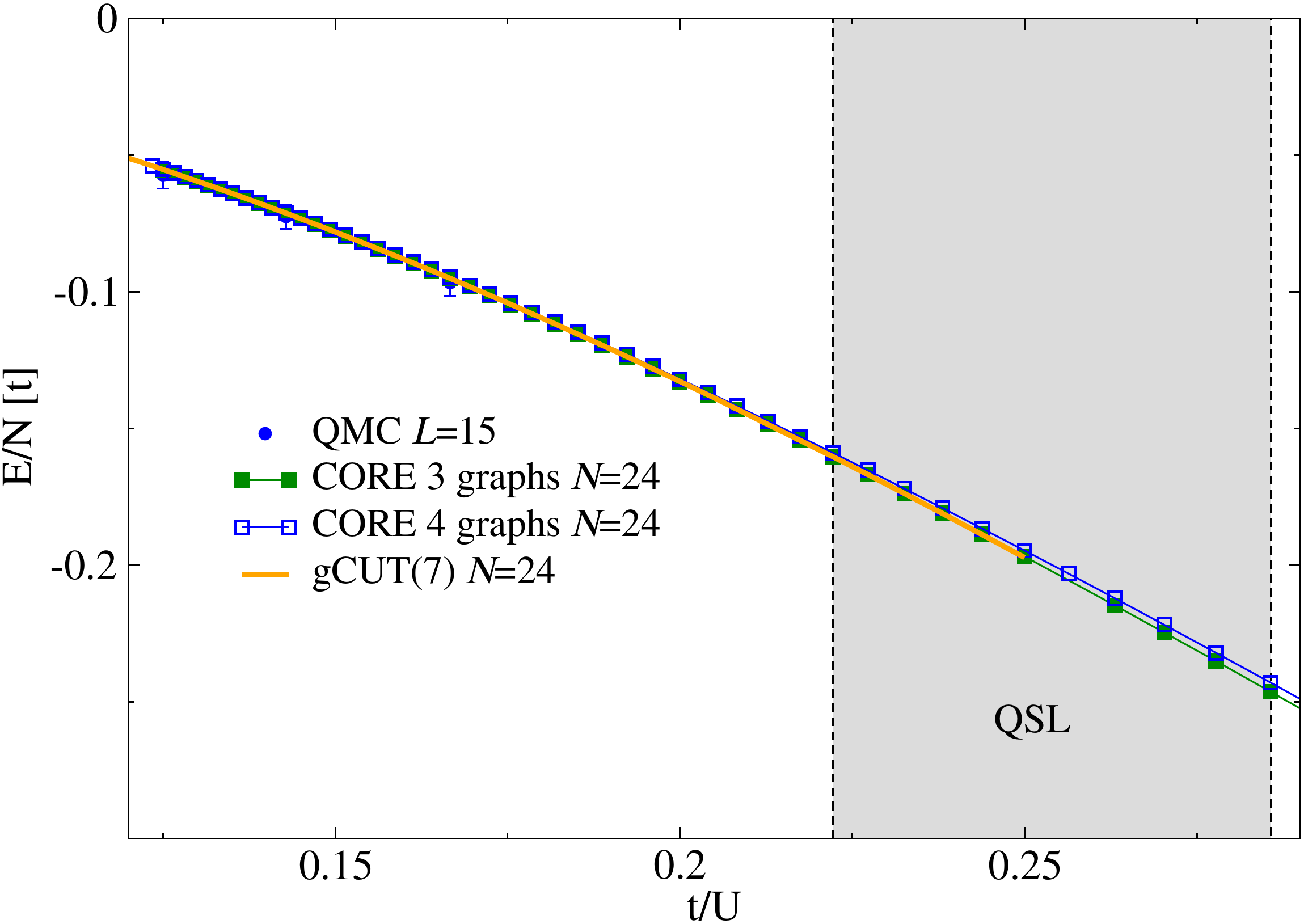}
\end{center}
\caption{(Color online) Ground-state energy per site as a function of $t/U$. A comparison is made to ED ($N=18$) and QMC ($N=72$)~\cite{Thank_Meng} data of the Hubbard model.} 
\label{fig:e0_vs_U}
\end{figure}

We start by comparing the ground-state energy per site as a function of $t/U$ which is displayed in Fig.~\ref{fig:e0_vs_U}. It can be clearly seen that all three different sets of graphs give a very similar energy
 per site in the full Mott phase which also agrees well with results for the Hubbard model. This is promising, but it is also a consequence of the fact that the ground-state energy per site is a rather unsensitive quantity.

We therefore turn to more sensitive quantities, namely the double occupancy per site and the derivative of the kinetic energy as discussed in the last subsection for the minimal effective model. Consequently, details of the considered effective spin models matter which is most clearly seen for the derivative of the kinetic energy. Interestingly and surprisingly, the agreement between results from the effective spin model and results for the Hubbard model gets worst at intermediate couplings. Indeed, already the minimal set of graphs plus the four-site star graph yields the wrong tendency. Furthermore, exact diagonalization of the full effective spin model from gCUT(7) does not even display the maximum in the derivative of the kinetic energy.
    
\begin{figure}
\begin{center}
 \includegraphics*[width=0.8\columnwidth]{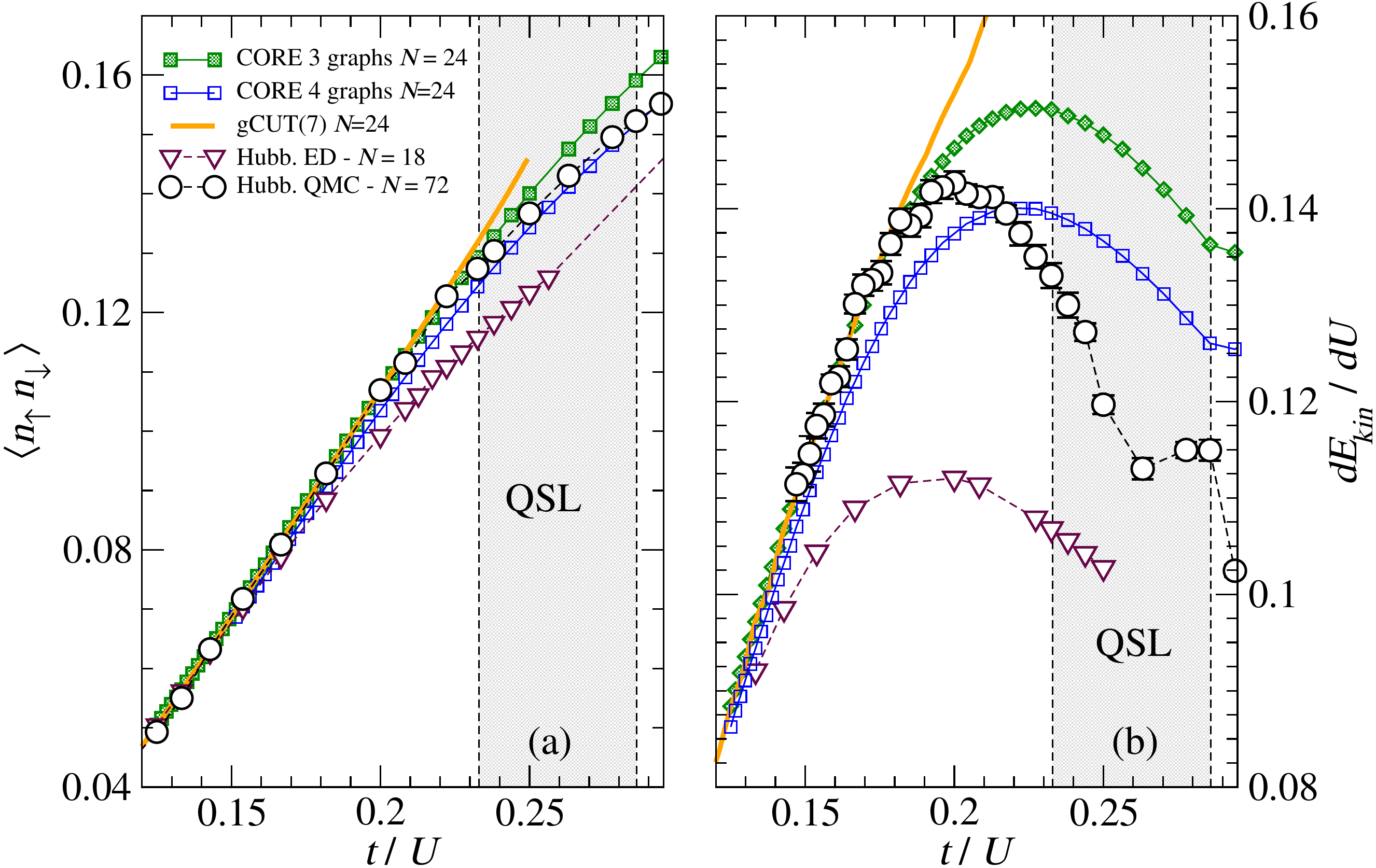}
\end{center}
\caption{(Color online) (a) Double occupancy per site obtained obtained with CORE effective model on finite clusters with  $N=18$ to $N=30$. (b) Derivative of the kinetic energy, as a function of $t/U$, obtained from the ground-state energy of the effective model [Eq.~(\ref{Ekin})] on the same clusters.  
In both cases, comparison is made to ED ($N=18$) and QMC ($N=72$)~\cite{Thank_Meng} data of the Hubbard model.} 
\label{fig:kin_vs_U_2}
\end{figure}

Altogether, we find that the inclusion of more graphs in our calculation gives a reduced agreement at intermediate couplings in contrast to our expectation. This trend is most probably neither a numerical problem nor a difference between CORE and gCUT, because both techniques agree quantitatively for both restricted set of graphs. In our opinion there are basically two possible scenarios. First, an effective spin model can still be reliably derived at intermediate couplings using a hexagon expansion. Indeed, the minimal model only focusing on the leading one-hexagon graphs gives very nice results. It might then be possible that the full graph decomposition used for the gCUT by including all graphs up to seven sites does not give improved results, because the relevant low-energy physics is contained in graphs with longer loops (these are exactly the multi-hexagon graphs). Second, the trend seen in the full graph decomposition by gCUT is a real effect, i.e. a controlled derivation of an effective low-energy spin model for the QSL at intermediate couplings remains a big challenge.

\section{Conclusions}
\label{Conclusions}

Summarizing, we have derived effective spin Hamiltonians describing the low-energy physics of the Hubbard model on the honeycomb lattice [Eq.~(\ref{Hubbard_model})]
at half-filling, that has been recently shown to display a QSL phase \cite{Meng10}. The effective models, obtained from the non-perturbative gCUT \cite{Yang11} and CORE
\cite{CORErefs,Capponi2004} methods, are in good mutual agreement and seemingly well converged for the intermediate values of $t/U$ yielding a QSL \cite{Meng10}.
We find that the effective next-NN two-spin frustrating exchange $J_2$ is not sizeable enough to destabilize N{\' e}el order \cite{Albuquerque11,Reuther11,Oitmaa11,Mosadeq11,
Mezzacapo12} and therefore cannot account for the existence of the QSL phase for Eq.~(\ref{Hubbard_model}). Instead, we find that the dominant sub-leading terms
in the effective model are six-spin exchanges (see Fig.~\ref{fig:couplings}), that may account for the emergence of the QSL behavior.

Additionally, we have taken some first steps in trying to characterize the so-obtained effective model, by performing exact diagonalizations on small clusters. While the
so-called Anderson tower of states is consistent with the occurrence of N{\' e}el order in the strongly interacting limit of $t/U \rightarrow 0$, the situation is less clear for
$t/U = 0.25$ (that according to Ref.~\cite{Meng10} yields a QSL), possibly an indication of a magnetically disordered state.

Naturally, it would be desirable to unambiguously show that the herein derived effective model displays a QSL as its ground-state and possesses a gap to spin excitations. However, the original results for the Hubbard model \cite{Meng10} indicate that such spin gap is very small and thus one presumably needs to consider clusters comprising
several hundred sites, so to surpass the spin correlation length and convincingly establish the nature of the ground-state. 

In face of this limitation, we can only conclude that the minimal effective spin model derived in the present work gives very promising results, but that a comprehensive characterization of its ground-state is still missing. Accordingly, we hope that our results may stimulate further work in trying to understand effects due to the six-spin interactions $L_1,L_2,\cdots L_5$ depicted in Fig.~\ref{fig:couplings}. Physically, one expects that these six-spin interactions frustrate the N\'eel order leading to the stabilization of a QSL. One possible strategy would be to deform the here obtained effective model and to study its phase diagram in an enlarged parameter space, in the hope that the QSL would be further stabilized and the spin correlation length would be reduced within reachable system sizes.

Moreover, as shown by Lieb~\cite{Lieb}, the ground-state of the
Hubbard model on any bipartite lattice satisfies the Marshall-Peierls
sign rule~\cite{Marshall1,Marshall2} \emph{exactly} for the singly
occupied configurations. By numerically diagonalizing the minimal
effective Hamiltonian derived from CORE, we have checked that this is
also the case for the spin model, which is a highly non-trivial result
since a simple $J_1-J_2$ spin model on the honeycomb lattice strongly
violates this rule in the intermediate region $0.2 < J_2/J_1 <
0.4$. While this property strongly constraints the nodes of the
wavefunction, it is however not clear if one could simulate the
effective spin model with a quantum Monte-Carlo algorithm
without minus-sign problem. Such an opportunity would allow to use
spin QMC techniques which are far more efficient that fermionic QMC
algorithm and thus could reach much larger system size. Clearly, this
point deserves future studies, especially since Sorella {\it et al.} have recently challenged the 
existence of the quantum spin liquid phase~\cite{Sorella}.

\ack
 K.P.S. acknowledges ESF and EuroHorcs for funding through his EURYI. A.~F.~A.~acknowledges financial support from Faperj (Brazil). 
S. C. acknowledges IUF for funding and KITP for hospitality. 
This research was supported in part by the National Science Foundation under Grant No. NSF PHY11-25915.
We acknowledge stimulating discussions with F.~Assaad, Z.~Y.~Meng, T.~C.~Lang, F.~Mila, A.~Muramatsu, A.~Paramekanti, S. Sorella and S.~Wessel.


\begin{thebibliography}{13}
\bibitem{gap_remark} For a fixed particle number, the threshold to particle-hole excitation is given by the optical gap that we have not computed. Instead, we give the one-particle gap energy scale which is expected to be of the same order of magnitude.
\end{thebibliography}
\end{document}